\documentclass{emulateapj}

\newcommand{\Msun}{\ensuremath{M_\odot}}
\newcommand       \kms		{\,{\rm km \,\, s}^{-1}}

\shorttitle{Origin of Young Stars in M31}
\shortauthors{Chang, Murray-Clay, Chiang, \& Quataert}

\begin{document}

\title{The Origin of the Young Stars in the Nucleus of M31}
\author{Philip Chang\altaffilmark{1,2}, Ruth
  Murray-Clay\altaffilmark{1}, Eugene Chiang\altaffilmark{1}, \&  Eliot Quataert\altaffilmark{1}} 
\altaffiltext{1} {Astronomy Department and Theoretical Astrophysics
  Center, 601 Campbell Hall, University of California, Berkeley, CA
  94720; pchang@astro.berkeley.edu, rmurray@astro.berkeley.edu,
  echiang@astro.berkeley.edu, eliot@astro.berkeley.edu}
\altaffiltext{2} {Miller Institute for Basic Research}

\begin{abstract}
  The triple nucleus of M31 consists of a population of old red stars
  in an eccentric disk (P1 and P2) and another population of younger A
  stars in a circular disk (P3) around M31's central supermassive
  black hole (SMBH).  We argue that P1 and P2 determine the maximal
  radial extent of the younger A star population and provide the gas
  that fueled the starburst that generated P3.  The eccentric stellar
  disk creates an $m=1$ non-axisymmetric perturbation to the
  potential.  This perturbed potential drives gas into the inner
  parsec around the SMBH, if the pattern speed of the eccentric
  stellar disk is $\Omega_p \lesssim 3-10\,{\rm km\,s^{-1}\,pc^{-1}}$.
  We show that stellar mass loss from P1 and P2 is sufficient to
  create a gravitationally unstable gaseous disk of $\sim 10^5\Msun$
  every $0.1-1$ Gyrs, consistent with the 200 Myr age of P3. Similar
  processes may act in other systems to produce very compact nuclear
  starbursts.
\end{abstract}


\section{Introduction}\label{sec:intro}

In balloon-borne experiments, Light et al. (1974) discovered that the
nucleus of M31 is asymmetric. Observations with the Hubble Space
Telescope (HST) resolved the nucleus into two components (Lauer
et al. 1993).  The two nuclei, denoted P1 and P2, have an angular
separation of 0''.5 ($\approx 2$ pc at the distance of M31; Bender et
al.  2005, hereafter B05).  P2 is located near the dynamical center
while P1, the brighter nucleus, is offset. P1 and P2 have a total
luminosity of $\approx 3\times 10^6\,{\rm L_{\odot}}$.  For a
mass-to-light ratio of $5.7$ appropriate for a bulge population
(Tremaine 1995, hereafter T95), the total stellar mass is $\approx
2\times 10^7\Msun$.

P1 and P2 are unlikely to be separate stellar clusters because the
time for dynamical friction to merge such clusters would be short
($\sim 10^6\,{\rm yrs}$ if P1 is a $\sim 10^7\,{\rm M_{\odot}}$
cluster; T95).  Instead, P1 and P2 are best modeled as a single
eccentric stellar disk, as originally proposed by T95 (see also Peiris
\& Tremaine 2003, hereafter PT03).  Disk stars slow near apocenter,
giving rise to P1.  Stars near pericenter, and at small disk radii, give
rise to P2.  The central supermassive black hole (SMBH) sits within
P2.

Spectroscopically, P1 and P2 are nearly identical, consistent with the
eccentric stellar disk model.  However, P2 is bluer.  Neito et al.
(1986) showed that P2 is brighter than P1 at 3750 \AA.  King et al.
(1995) showed that P2 is brighter than P1 in the ultraviolet.  Though
this difference was initially attributed to an active galactic nucleus
(AGN), recent HST spectroscopy of the nuclear region of M31 has
uncovered a younger population of 200-Myr old A stars, embedded in P2
(B05).  This population, named P3, appears to be a disk of stars with
a mass of $\sim 4200\Msun$ and a maximal radial extent of $\approx 1$
pc (B05; Bender priv.  communication) that surrounds the central SMBH
and lies in the same plane as the P1/P2 disk. Using the line-of-sight
velocities ($\approx 1000\,{\rm km\,s}^{-1}$) measured for P3, B05
estimate that the mass of the SMBH in M31 is $1.1-2.1\times
10^8\,\Msun$.


P3 is a stellar population distinct from P1 and P2 (B05).  P3 is
composed of A stars while P1 and P2 are typical old red bulge stars.
The relative youth of these A stars and their proximity to the central
SMBH make P3 analogous to the young stars in our Galactic Center (GC).
Like the young stars in the GC, P3 must either form in situ or migrate
in from larger radii.  Migration is less likely in M31 than in the GC
as the progenitor cluster would be disrupted at greater distances from
the more massive SMBH in M31.  In situ formation more naturally
explains the masses of these central star-forming regions through
Toomre stability arguments (see \S~3).  However, it is less clear what
sets the radial extents ($r \lesssim 1$ pc for M31; $r \lesssim 0.4$
pc for the GC) and ages ($\approx 200$ Myr for M31; $\approx 10$ Myr
for the GC) of these nuclear starbursts.

In this paper, we address these questions by demonstrating that the
eccentric stellar disk of M31 fixes both the radial extent and the
timescale for the starburst that generated P3.  In
\S~\ref{sec:orbits}, we argue that the non-axisymmetric potential from
the eccentric stellar disk limits non-intersecting gas orbits to a
limited family around the central SMBH.  The physics is similiar to
what sets the maximum sizes of accretion disks in Roche-lobe filling
binaries (Paczynski 1977; Papaloizou \& Pringle 1977).  We present
numerical and analytic calculations describing how non-intersecting
gas orbits are only allowed for $r\lesssim 1$ pc if the pattern speed of the
P1/P2 disk is $\lesssim 3-10\,{\rm km\,s^{-1}\,pc^{-1}}$.  This
naturally explains the size of P3.  We then argue in \S~\ref{sec:gas}
that stellar mass loss from the P1/P2 disk is sufficient to supply the
gas needed to form P3.  We estimate the mass required to trigger a
starburst and the timescale to build up that mass, and show that these
are consistent with the mass and age of P3.  Finally, we conclude in
\S~\ref{sec:conclusions}, drawing attention to a few predictions of
our model and arguing that this mechanism may be common in galactic
nuclei.

\section{Closed Gas Orbits in an Eccentric Stellar Disk}\label{sec:orbits}

In the limit that gas has zero pressure, gas follows test particle
orbits that are simply closed and non-crossing (Paczynski 1977).  Gas
not in these orbits will shock against gas in neighboring orbits
(i.e., crossing) or itself (i.e., not simply closed).  These shocks
dissipate orbital energy, driving gas to lower energy orbits.
Paczynski (1977) applied this principle to solve for the maximum size
of a gaseous accretion disk in a Roche-filling binary star system.
Test particle orbits that are close to the accretor, where the
potential is nearly Keplerian, can be nearly circular and
non-crossing.  Farther from the accretor, the non-axisymmetric
component of the potential (due to the donor star) becomes larger
until test particle orbits are forced to cross either their neighbors
or themselves.  Therefore there exists a maximum radius for gas orbits
in the vicinity of the accretor: the tidal truncation radius, $R_{\rm
t}$ (Papaloizou \& Pringle 1977).  Gas outside $R_{\rm t}$ will be
driven toward it through dissipative processes, while gas inside
$R_{\rm t}$ will occupy the allowed orbits, forming a gaseous
accretion disk.  Paczynski (1977) showed that only one family of
orbits is possible for any given binary system with a specified mass
ratio. These results were later confirmed by numerical simulations of
close binary systems by Ichikawa \& Osaki (1994).

By analogy, P1 and P2 add a non-axisymmetric component to the point
mass potential of the central SMBH in M31. Thus, there should also be an
$R_{\rm t}$ inside of which a gaseous accretion disk can exist around
the SMBH.  While this situation is similar to that of a close binary
star system, there are two differences.  First, the perturbation to
the potential in M31 is given by the eccentric stellar disk.  Second,
whereas the pattern speed of the perturbation potential in a binary
star system is prescribed by the binary orbital frequency, the pattern
speed of the P1/P2 disk (i.e., its apsidal precession frequency) is
uncertain (see Appendix A for estimates in the literature).

Since the gas mass required to form P3 ($M_{\rm gas} \lesssim
10^5\Msun$; see \S~\ref{sec:gas}) is much smaller than the mass of the
P1/P2 disk and the central SMBH, and since we assume that Toomre's
$Q>1$ for the present discussion (\S~\ref{sec:gas}), we neglect gas
self-gravity.  Moreover, since the characteristic temperature of the
gas is $T \sim 30$ K (\S~\ref{sec:gas}), the thermal energy of a gas
particle, $\sim kT$, where $k$ is Boltzmann's constant, is much
smaller than the particle's gravitational energy due to the P1/P2
disk, $\sim G M_D\mu/R_D$. Here $G$ is the gravitational constant,
$M_D \approx 2\times 10^7\Msun$ and $R_D \sim 1\,{\rm pc}$ are the
mass and characteristic size of the disk, respectively, and $\mu$ is
the mean molecular weight of the gas.  Therefore, as Paczynski (1977)
originally envisioned, gas orbits can be computed in the zero pressure
limit.

To calculate $R_{\rm t}$, we look for simply closed and non-crossing
orbits in the combined potential of a central SMBH and an eccentric,
precessing stellar disk.  All orbits are computed in the $(x,y)$ plane
of the eccentric stellar disk.  We orient P1 and P2 such that P1 is
centered at $x = -3$ pc (which corresponds to the 2 pc separation
projected on the sky).  The SMBH is fixed at the origin.  We assume
that the P1/P2 disk precesses rigidly counterclockwise with pattern
speed $\Omega_p$ about the center of mass, located at $(x_{\rm cm},
0)$.  In the rotating frame, the equations of motion are
\begin{eqnarray}\label{eq:eom}
  \ddot{x} &=& -\frac {d\Phi} {dx} + \Omega_p^2 \left(x - x_{\rm cm}\right) + 2\Omega_p \dot{y} \\
  \ddot{y} &=& -\frac {d\Phi} {dy} + \Omega_p^2 y - 2\Omega_p \dot{x},
\end{eqnarray}
where the potential $\Phi$ is given by
\begin{equation}
\Phi(x,y) = -\frac {GM_{\rm BH}} r + \Phi_{\rm D}(x,y),
\end{equation}
where $M_{\rm BH}$ is the mass of the central SMBH, $r = \sqrt{x^2 +
  y^2}$, and $\Phi_{\rm D}$ is the potential due to the eccentric
disk:
\begin{equation}\label{eq:disk potential}
\Phi_{\rm D}(x, y) = -G\int dx' dy' \frac {\Sigma(x', y')}{|r - r'| + h} 
\end{equation}
where $|r - r'| = \sqrt{(x - x')^2 + (y-y')^2}$, $\Sigma$ is the
stellar mass surface density, and $h$ is the softening length to
account for the finite thickness of the disk.  For now, we take a
fixed softening length of $h = 0.1$ pc to cleanly demonstrate how the
P1/P2 disk can tidally truncates a gaseous disk.  The P1/P2 disk may
be substantially thicker, however.  At the end of this section, we
discuss various choices for $h$ and their effects on our results.

For $\Sigma$, we use the fit to
the light distribution from PT03 and a mass-to-light ratio of 5.7
(T95; PT03).  We focus on PT03's non-aligned model as it fits the
light distribution and kinematic data better for P1 and P2.  In
addition, we ignore the contribution to the potential from the local
bulge because its mass is only 10\% of that of the disk ($\lesssim
10^6\Msun$; PT03). We refer the interested reader to Appendix
\ref{sec:PT} for more details.



The strength of the non-axisymmetric component relative to the
axisymmetic component depends on the mass of the P1/P2 disk relative
to that of the SMBH, $M_D/M_{\rm BH}$.  PT03 give a stellar mass of
$\approx 2\times 10^7\,\Msun$ for a mass-to-light ratio appropriate
for the bulge and fit a SMBH mass of $10^8\Msun$.  Salow \& Statler
(2001, 2004) construct a series of self-consistent dynamical models,
which give stellar masses of $1-2\times 10^7\Msun$ and a SMBH mass of
$\approx 5\times 10^7\Msun$.  B05 give a SMBH mass of $1.4\times
10^8\Msun$ ($1.1-2.1 \times 10^8\Msun$ at the 1$\sigma$ error level)
based on the dynamics of P3.  The range of mass ratios, $M_D/M_{\rm
  BH}$, from these different authors is $\approx 0.1-0.3$.  We adopt a
fiducial ratio of $M_D/M_{\rm BH}=0.1$ motivated by the stellar disk
mass from PT03 and the SMBH mass from B05; in this case $x_{\rm cm} =
-0.07$ pc.

\begin{figure}
\plotone{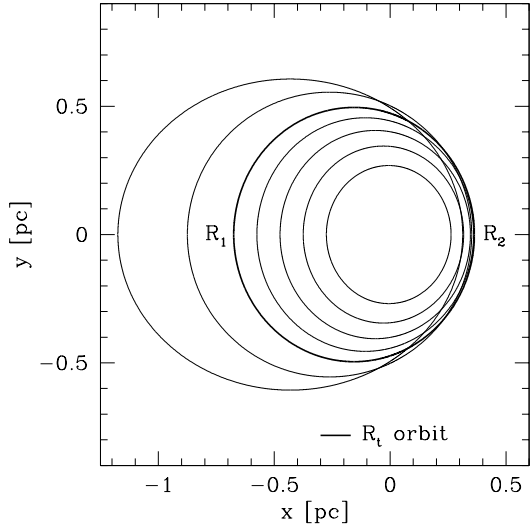}
\caption{Gas orbits for $\Omega_p = 3\,{\rm km\,s^{-1}pc^{-1}}$,
  $M_{\rm BH} = 10^8\Msun$, $M_D/M_{\rm BH} = 0.1$, and a softening
  length of $h=0.1$ pc. The largest possible orbit, denoted $R_{\rm
    t}$, is shown with the thick solid line.  The outermost two orbits
  have their closest approach inside of the $R_{\rm t}$ orbit (they
  also cross each other).  Gas in orbits exterior to the $R_{\rm t}$
  orbit will shock and be driven to it.  Note that the $R_{\rm t}$
  orbit is very eccentric.}
\label{fig:3km}
\end{figure}

We compute test particle (gas) orbits using equations (\ref{eq:eom})
and (2).
The orbit starts at $x = -R_1$ with a velocity purely in the $-y$
direction ($\dot{x} = 0$); see Figure \ref{fig:3km}.  Stars rotate
about the SMBH in the counterclockwise direction.  We take the gas
rotation and the P1/P2 disk pattern speed to also be in the
counterclockwise directions.  For each computed orbit, we check to see
if the ending position and velocity are the same as the beginning
position and velocity. We vary the initial velocity $\dot{y}$ until
this condition is met. We define $R_2$ as the position where the orbit
crosses the positive $x$-axis.  Repeating this calculation over a
range of $R_1$, we find a family of simply closed orbits for a given
pattern speed $\Omega_p$.

Figure \ref{fig:3km} shows such simply closed orbits for $M_D/M_{\rm
  BH} = 0.1$, $M_{\rm BH} = 10^8\Msun$, and $\Omega_p = 3\,{\rm
  km\,s^{-1}pc^{-1}}$.  At such a low pattern speed, gas orbits are
restricted to lie inside a maximal orbit (thick solid line), which we
denote the $R_{\rm t}$ orbit.  Gas outside this orbit will cross
the $R_{\rm t}$ orbit (as in the outermost two orbits in Fig.
\ref{fig:3km}) or be self-crossing.

\begin{figure}
\epsscale{1.0} 
\plotone{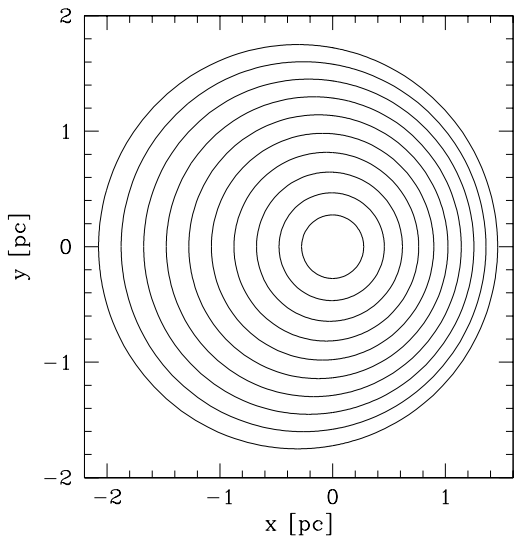}
\caption{Same as Figure \ref{fig:3km}, except $\Omega_p = 30\,{\rm
    km\,s^{-1}pc^{-1}}$.  Unlike in Figure \ref{fig:3km}, there is no
  $R_{\rm t}$ orbit.  Gas finds non-crossing orbits at all
  radii.}
\label{fig:30km}
\end{figure}

Figure \ref{fig:30km} shows orbits for a higher pattern speed
$\Omega_p = 30\,{\rm km\,s^{-1}pc^{-1}}$.  These orbits form
 non-crossing orbits spanning the entire disk.  Figures
\ref{fig:3km} and \ref{fig:30km} illustrate that the nature of gas
orbits qualitatively changes when going from low to high pattern
speeds. Gas occupies restricted orbits at low pattern speeds, while at
higher pattern speeds, gas can span the entire disk.

\begin{figure}
\plotone{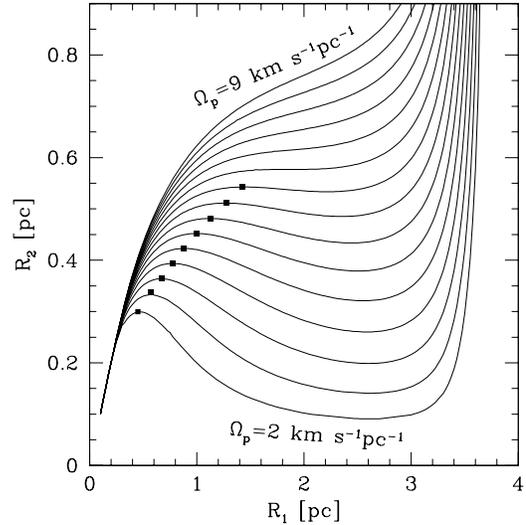}
\caption{$R_2$ (periapse distance) as a function of $R_1$ (apoapse
  distance) for $\Omega_p = 2-9\,{\rm km\,s^{-1}pc^{-1}}$ in
  $0.5\,{\rm km\,s^{-1}pc^{-1}}$ intervals, for $M_D/M_{\rm BH} =
  0.1$, $M_{\rm BH} = 10^8\Msun$, and $h=0.1$ pc.  For small
  $\Omega_p$, $R_2$ obtains a local maximum, which we mark with a
  solid black square.  For $\Omega_p > 6\,{\rm km\,s^{-1}\,pc^{-1}}$,
  this local maximum does not exist.}
\label{fig:r2}
\end{figure}

We plot $R_2$ as a function of $R_1$ for different values of
$\Omega_p$ in Figure \ref{fig:r2}.  For small $\Omega_p$, $R_2$ has a
local maximum, $R_{\rm 2,max}$ (marked with solid black squares in
Fig.  \ref{fig:r2}).  We denote the $R_1$ for which this occurs as
$R_{\rm 1,max}$.  Beyond $R_1 \approx 3$ pc, $R_2$ increases with
$R_1$ for all $\Omega_p$.



\begin{figure}
\plotone{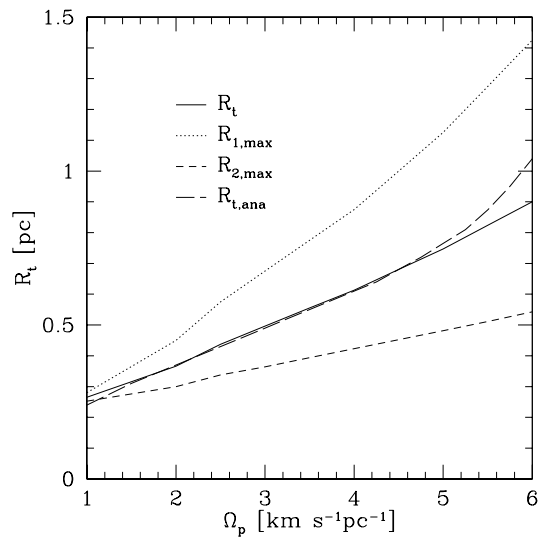}
\caption{$R_{\rm t}$ (solid line) as a function of $\Omega_p$ for
  $M_D/M_{\rm BH} = 0.1$, $M_{\rm BH} = 10^8\Msun$, and $h=0.1$ pc.  We also plot
  $R_{\rm 1, max}$ (dotted line), $R_{\rm 2,max}$ (short dashed line), and
  $R_{\rm t, ana}$ (long dashed line) as defined by the crossing condition
  (eq.[\ref{eq:crossing condition}]).  For $\Omega_p \gtrsim 6\,{\rm
    km\,s^{-1}\,pc^{-1}}$, gas orbits can persist at all radii.}
\label{fig:rmax}
\end{figure}

We define the tidal truncation radius as the angle-averaged radius of
the maximal non-intersecting orbit: $R_{\rm t} = (2\pi)^{-1}\int
R(\theta)d\theta$.  Figure \ref{fig:rmax} shows $R_{\rm t}$ as a
function of $\Omega_p$.  As $\Omega_p$ increases, $R_{\rm t}$
increases.  The tidal truncation radius, $R_{\rm t}$, is $\lesssim 1$
pc (similar to the observed maximal radial extent of P3) when
$\Omega_{p} \lesssim 6\,{\rm km\,s^{-1}\,pc^{-1}}$ for $M_D/M_{\rm BH}
= 0.1$ and $M_{\rm BH} = 10^8\Msun$.  For larger pattern speeds,
$R_{\rm t}$ does not exist and a gaseous disk can span all radii.



\begin{figure}
\epsscale{1.0} 
\plotone{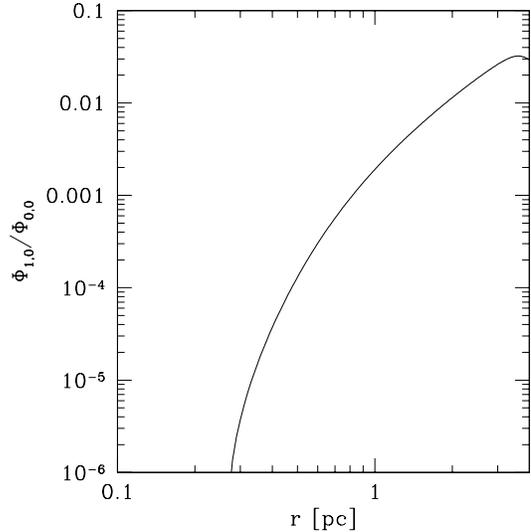}
\caption{Ratio of $\Phi_{1,0}$ ($m=1$ component of the potential) to
  $\Phi_{0,0}$ (axisymmetric component of the potential) as a function
  of $r$ for $M_D/M_{\rm BH} = 0.1$.}
\label{fig:fourier}
\end{figure}

The dependence of $R_{\rm t}$ on $\Omega_p$, $M_D$, and $M_{\rm BH}$
can be derived using perturbation theory (e.g., Papaloizou \& Pringle
1977; Binney \& Tremaine 1987).  In cylindrical coordinates ($r$,
$\phi$), the equations of motion in the frame rotating at $\Omega_p$
are
\begin{eqnarray}
  \ddot{r} &=& -\frac {d\Phi}{dr} + \frac {l^2} {r^3} + \frac {2\Omega_p l}{r} + \Omega_p^2 r  \\
  \dot{l} &=& -\frac {d\Phi}{d\phi} - 2\Omega_p r \dot{r},
\end{eqnarray}
where $l = r^2\dot{\phi}$ is the specific angular
momentum.\footnote{We assume in the following that the disk precession
  axis passes through the origin (SMBH).} Following the derivation in
Binney \& Tremaine (1987), we take $\Phi\rightarrow \Phi_0(r) +
\sum_{m=1}^{\infty}\Phi_m(r,\phi)$, $r \rightarrow r + \delta r$ and $\phi
\rightarrow \phi + \delta\phi$ in the epicyclic approximation, where
$\Phi_m = \Phi_{m,0}(r)\cos\left[m(\Omega - \Omega_p) t\right]$ are
the Fourier components of the potential and $\Omega =
\sqrt{r^{-1}d\Phi_0/dr}$ is the orbital frequency at the guiding center, $r$.
If the axisymmetric component ($\Phi_0$) is dominant, the solution
for $\delta r$ is (Binney \& Tremaine 1987, their eq.[3.119ab])
\begin{eqnarray} \label{eq:delta_r} 
  \delta r = -\frac
  {\cos\left(m(\Omega - \Omega_p)t\right)} {\kappa^2 -
    m^2(\Omega - \Omega_p)^2} \left(\frac {d} {dr} + \frac
    {2\Omega} {\left(\Omega-\Omega_p\right)r}\right)\nonumber \\\Phi_{m,0}(r)
  + C\cos\left(\kappa t + \psi\right),
\end{eqnarray}
where $\kappa^2 = d^2\Phi_0/dr^2 + 3\Omega^2$ is the square of the
epicyclic frequency, $C$ and $\psi$ are constants, and we assume that
the perturbation is dominated by a single mode $m$.  Note that simply
closed orbits correspond to $C=0$.  For the eccentric stellar disk,
the dominant Fourier mode is $m=1$, whose amplitude we numerically
calculate and plot in Figure \ref{fig:fourier}.


Orbits first cross their neighbors at pericenter (i.e., $\delta r <
0$; see Figure \ref{fig:3km}) when
\begin{equation}\label{eq:crossing condition}
  \frac {d\delta r} {dr} < -1;
\end{equation}
i.e., when the epicyclic amplitude grows faster than the size of the
guiding center orbit (Papaloizou and Pringle 1977).  From our
numerically calculated $\Phi_{1,0}$, we evaluate ${d\delta r}/{dr}$ as
a function of $r$ and determine an analytic tidal truncation radius,
$R_{\rm t, ana}$, where ${d\delta r}/{dr}$ first equals $-1$. We plot
$R_{\rm t, ana}$ in Figure \ref{fig:rmax} to compare to the
numerically calculated $R_{\rm t}$.  The agreement is good.


Since the potential is nearly Keplerian ($\kappa \approx \Omega$) and
since $\Omega_p/\Omega \ll 1$, we expand equation (\ref{eq:delta_r})
to first order in $\Omega_p/\Omega$ and $\Phi_{1}/\Phi_0$.
For $d/dr \sim 1/r$, the condition for first orbit crossing
(eq.[\ref{eq:crossing condition}]) becomes
\begin{equation}\label{eq:breakdown}
  \frac {|\delta r|} {r} \sim \frac {\Omega} {\Omega_p} \frac {\Phi_1}{\Phi_0} \sim 1,
\end{equation}
where we drop numerical factors.  Equation (\ref{eq:breakdown})
indicates that for the fixed $R_{\rm t}$ orbit, $\Omega_p \propto \Omega
{\Phi_1}/{\Phi_0}$.  Since ${\Phi_1}/{\Phi_0}$ scales as $M_D/M_{\rm
  BH}$ and $\Omega$ scales as $M_{\rm BH}^{1/2}$, we find:
\begin{equation}\label{eq:omega_scaling}
  \Omega_p \propto M_{\rm BH}^{-1/2}M_D.
\end{equation}
We demonstrated earlier (see Figure \ref{fig:rmax}) that gas orbits
are limited to an inner disk similar to the maximal radial extent of
P3 (i.e., $R_{\rm t} < 1$ pc) if the pattern speed $\Omega_p \lesssim
6\,{\rm km\,s^{-1}\,pc^{-1}}$ for $M_{\rm BH} = 10^8\Msun$ and
$M_D/M_{\rm BH} = 0.1$.  Inserting the scalings from
(\ref{eq:omega_scaling}), we rescale this pattern speed to be
\begin{equation}\label{eq:pattern speed}
  \Omega_p \lesssim 6\,\left(\frac {M_{\rm BH}}{10^8\Msun}\right)^{-1/2}\left(\frac {M_D}{10^7\Msun}\right)\,{\rm km\,s^{-1}\,pc^{-1}}.
\end{equation}
We have confirmed these scalings numerically.  For M31, if $M_{D}
\approx 2\times 10^7\Msun$ (PT03) and $M_{\rm BH} \approx 1.4 \times
10^8\Msun$ (B05), then equation (\ref{eq:pattern speed}) gives
$\Omega_p \lesssim 10 \,{\rm km\,s^{-1}\,pc^{-1}}$.

We now return to the issue of the thickness of the P1/P2 disk and its
impact on $R_{\rm t}$.  The non-aligned model of PT03, which we use
for the surface density profile, uses $h/r \approx 0.4$. On the other
hand, the disk models of Salow and Statler (2001, 2004) are razor
thin.  Bacon et al. (2001) also advocate a cold thin disk ($h/r \sim
0.1$) to model P1 and P2.  Toomre stability arguments give a
minimum $h/r$ of $\approx 0.1$ (PT03; see also Bacon et al. 2001).  If
the P1/P2 disk has persists for $10^{10}$ years, two-body relaxation
gives a minimum $h/r \approx 0.2$ (PT03; T95).

\begin{figure}
\plotone{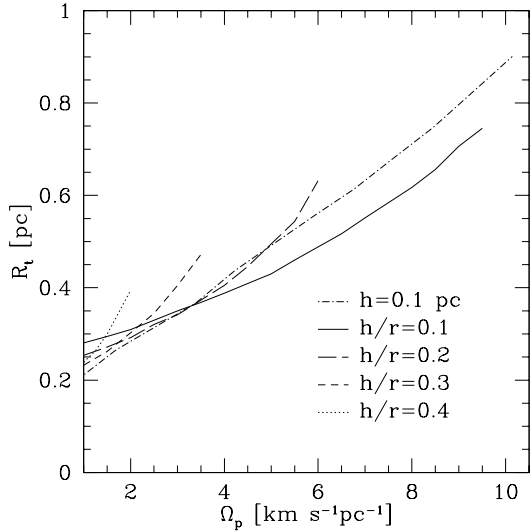}
\caption{$R_{\rm t}$ as a function of $\Omega_p$ for parameters
  appropriate to M31 ($M_D = 2\times 10^7\Msun$ and $M_{\rm BH} = 1.4
  \times 10^8\Msun$) for various softening parameters of $h/r =$ 0.1
  (solid line), 0.2 (long-dashed line), 0.3 (short-dashed line), and
  0.4 (dotted line).  We also show the softening length, $h=0.1$ pc,
  (dashed-dotted line) for comparison.}
\label{fig:rtidm31}
\end{figure}

How does varying the softening length affect our results?  In Figure
\ref{fig:rtidm31}, we plot $R_{\rm t}$ as a function of $\Omega_p$ for
various softening parameters of $h/r = 0.1 - 0.4$, taking $M_D =
2\times 10^7\Msun$ and $M_{\rm BH} = 1.4 \times 10^8\Msun$ as is
appropriate for M31.  The results for $h=0.1$ pc, which we have
focused on thus far, differ from those for $h/r = 0.1$ by $\approx
10\%$.
However, as Figure \ref{fig:rtidm31} shows, thicker disks show more
substantial differences.  As we increase the softening parameter from
$h/r = 0.1$ to $0.4$, the maximum $\Omega_p$ for which $R_{\rm t}$
exists decreases, down to $\approx 2\,{\rm
  km\,s^{-1}pc^{-1}}$.  In addition, the maximum $R_{\rm t}$ also
decreases, down to $\approx 0.4$ pc.
To produce $R_{\rm t} \lesssim 1$ pc, similar to the observed maximal
radius ($\sim 1$ pc) of the P3 disk (B05), we require $\Omega_p
\lesssim 3-10\,{\rm km\,s^{-1}pc^{-1}}$ for $h/r = 0.1 - 0.3$.

Finally, we estimate the time, $t_{\rm flow}$, for gas to flow to $R_{\rm
  t}$ from larger radii.  Compared to the axisymmetric component of
the potential, the $m=1$ component is smaller by of order
$10^{-3}-10^{-2}$ for $r\gtrsim 1$ pc (see Fig.  \ref{fig:fourier}).
We expect the epicyclic velocity to be $\delta v \sim
\sqrt{\Phi_1/\Phi_0} v_{\rm orb} \approx 0.1 v_{\rm orb} \approx
70\,{\rm km\,s^{-1}}$ at 1 pc, where $v_{\rm orb}$ is the orbital
velocity.  This is much greater than the gas sound speed, $c_s \sim
0.3-1\,{\rm km\,s}^{-1}$ at $T\sim 10-100$ K.  Gas streams cross at
supersonic velocities, shock, and dissipate $\Phi_1/\Phi_0$ of their
orbital energy per dynamical time, $t_{\rm dyn} = \Omega^{-1}$.  Thus
the inflow time $t_{\rm flow} \sim t_{\rm dyn}
\left(\Phi_1/\Phi_0\right)^{-1}$, which is a few hundred to a few
thousand dynamical times.  At $r\sim 1$ pc, $t_{\rm flow} \sim
10^5-10^6$ yrs.

\section{Stellar Mass Loss as the Origin of the Gas that Formed P3}\label{sec:gas}

In \S~2 we argued that gas at $r \gtrsim R_{\rm t}$ is forced down to
$r \approx R_{\rm t}$ on short timescales, $t_{\rm flow} \sim
10^6\,{\rm yrs}$. Supplied with sufficient mass, the gas disk at
$\approx R_{\rm t}$ eventually becomes gravitationally unstable,
forming stars and giving rise to P3.  We now explore these questions
of gas supply and gravitational instability.

There are a number of potential sources of gas. For example, molecular
clouds can be gravitationally scattered onto low angular momentum
orbits. The rate of such gas supply is difficult to quantify, however.
Here we focus instead on mass supply due to winds from stars in the
P1/P2 disk.  This source of mass is unavoidable, existing even in the
absence of external sources.  We show below that mass loss from the
P1/P2 disk is sufficient to trigger a starburst having the right
properties to explain P3.



The P1/P2 disk consists of $\approx 2\times 10^7\Msun$ of stars with
an age of $\sim 10^{10}$ yrs (B05, PT03).  We compute the stellar mass
loss rate from Starburst99 (Leitherer et al. 1999; Vazquez \&
Leitherer 2005), using the Padova tracks with asymptotic giant branch
(AGB) stars.  A $\sim 10^{10}$ yr old star cluster of mass $2\times
10^7\,\Msun$ loses mass via stellar winds at a rate $\dot{M}_* \approx
3\times 10^{-5} - 3\times 10^{-4}\Msun\,{\rm yr^{-1}}$.  The mass loss
is primarily due to winds from red giants and thermally pulsating AGB
stars.  There are uncertainties in these mass loss rates due to
uncertainties in the properties of the thermally pulsating AGB stars.
The stellar winds which dominate the mass loss have velocities
($5-25\,{\rm km\,s}^{-1}$ for AGB stars; Lamers \& Cassinelli 1999)
much lower than the orbital velocity $v_{\rm orb}\sim 700\,{\rm
  km\,s}^{-1}$.  Hence the winds are bound to the nuclear region.  The
winds have typical relative velocities of $\sim (h/r) v_{\rm orb} \sim
200 \kms$, where $h \sim 0.3 \, r$ is the thickness of the P1/P2 disk.
The stellar winds thus shock, reaching temperatures of $\sim 10^6 \,
[(h/r)/0.3]^2$ K.  The fate of the shocked stellar wind material
depends on the competition between heating and cooling.  For rapid
cooling, the gas will collapse to form a thin disk.  For slow cooling,
it will heat up because of viscous stresses and form a geometrically
thick, radiatively inefficient accretion flow. 
The gas heats up on a characteristic timescale $t_{\rm heat} \approx
\alpha^{-1} \Omega^{-1}$, where $\alpha$ is the dimensionless
viscosity.\footnote{The heating from the gravitational torques exerted
  by the eccentric stellar disk is small compared to that from
  $\alpha$-viscosity for an initially thick disk.}  The cooling time
is given by $t_{\rm cool} \approx 3k T/2n \Lambda(T)$ where
$\Lambda(T)$ is the cooling function for an optically thin thermal
plasma. The density of stellar wind material that accumulates on a
timescale $t_{\rm heat}$ is $n \approx \dot{M}_* t_{\rm heat}/(2 \pi
r^3 [h/r] \mu)$.  If $t_{\rm cool} \lesssim t_{\rm heat}$ for gas at
this density, it will cool and collapse to form a thin disk.
This condition can be written as a constraint on the stellar wind mass
loss rate
\begin{equation}
\dot {M}_* \gtrsim \dot M_{\rm *, crit} \approx {3 \pi r \alpha^2 [h/r]^3
\mu^2 v_{\rm orb}^4 \over 5 \Lambda(T)}. \label{mdotcrit}
\end{equation}
Taking $\alpha = 0.1$, $M_{\rm BH} = 10^8 M_\odot$ and $r = 2$ pc as
fiducial numbers for M31, we find that $\dot M_{\rm *, crit} \approx 4
\times 10^{-7} - 6 \times 10^{-6} \, {\rm M_\odot \, yr^{-1}}$
for $h/r = 0.2-0.4$.  Since $\dot {M}_{*} \approx 3\times 10^{-5} - 3\times
10^{-4}\Msun\,{\rm yr^{-1}}$ for the stars in the P1/P2 disk, we conclude that 
stellar winds from P1/P2 will likely cool and collect in a
geometrically thin disk.\footnote{Our critical $\dot M_{\rm *,crit}$ is a factor of
$\gtrsim 20$ smaller than that usually estimated for the transition
from a thin to thick disk (e.g., Fig. 3 of Menou, Narayan, \& Lasota
1999).  The latter calculations assume $h \approx r$, i.e., that a
thick disk is already established, while in our problem the stellar
winds are initially confined to a region with $h \approx 0.3 \, r$.
Smaller $h/r$ increases $n$ and decreases $T$ at fixed $\dot M_*$, thus
significantly decreasing the cooling time of the gas.}



Cooled gas accumulates at $R_{\rm t} \sim 1$ pc until it either
accretes onto the SMBH or becomes gravitationally unstable and
fragments into stars (see related arguments of Nayakshin 2006 and
Levin 2007).  For a disk to fragment into stars, numerical studies
suggest that two conditions must be met (Gammie 2001; see also Shlosman \& Begelman 1989):
\begin{eqnarray}\label{eq:Toomre}
  Q &\lesssim& 1, \\ \label{eq:cooling}
  t_{\rm cool} &\lesssim& 3 t_{\rm dyn},
\end{eqnarray}
where $Q = c_s \kappa/\pi G\Sigma_g$ is the Toomre parameter, $c_s =
\sqrt{kT/\mu}$ is the gas sound speed, $\Sigma_g$ is the gas surface
density of the disk, and $t_{\rm dyn} = \Omega^{-1}$ is the local
dynamical time.  The radiative cooling time, $t_{\rm cool}$, is given
by
\begin{equation}\label{eq:coolingtime}
  \frac {\Sigma_g kT}{\mu t_{\rm cool}} \sim \sigma_B T^4
  \left\{\begin{array}{ll} 
      \tau_{\rm IR} & \tau_{\rm IR} \ll 1 \\
      \tau_{\rm IR}^{-1} & \tau_{\rm IR} \gg 1
    \end{array}
  \right.,
\end{equation}
where $\tau_{\rm IR} = \kappa_{\rm IR}\Sigma_g/2$ is the infrared (IR)
vertical optical depth of the gas disk, $T$ is the midplane
temperature of the disk, and $\kappa_{\rm IR}$ is the corresponding
opacity.  The first condition (eq. [\ref{eq:Toomre}]) is that gas
self-gravity must overcome rotation and gas pressure.  The second
condition (eq.  [\ref{eq:cooling}]) is that cooling is sufficiently
rapid to overcome heating due to gravitationally induced turbulence
(``gravitoturbulence''; e.g., Shlosman et al. 1989; Gammie 2001). If
equation (\ref{eq:Toomre}) is satisfied, but equation
(\ref{eq:cooling}) is not, then the disk enters a gravitoturbulent
state and accretes, but does not fragment or form stars.

The ability of the gas disk to fragment into stars thus depends on the
heating and cooling of the gas.  We consider two possibilities for the
heating: external heating by stellar irradiation from the P1/P2 disk
and intrinsic heating by gravitoturbulence.  We take the gas to cool
radiatively by emission from dust grains.\footnote{Gas and dust are
  thermally well coupled by gas-dust collisions.  A $\sim 0.1\micron$
  grain equilibrates with surrounding gas in $\sim 10^6\,{\rm s}$ at
  the gas densities $n \sim 8\times 10^8\,{\rm cm}^{-3}$ and
  temperatures $T\sim 30$ K characterizing $Q \sim 1$ disks.}
When the gas is externally heated by starlight, $Q>1$ initially for a
sufficiently low mass disk.  The disk mass grows from stellar winds
until $Q\sim 1$, when it becomes gravitationally unstable.  If
equation (\ref{eq:cooling}) is also satisfied, then the disk
fragments.

When external sources of heat are negligible, gravitoturbulence tends
to maintain the disk in a marginally stable state with $Q\sim 1$
(Gammie 2001).  Initially, the disk does not fragment because the
cooling time is long under these conditions.  As the mass in the disk
increases from stellar winds, the cooling time decreases relative to
the orbital period, and eventually fragmentation occurs.  Whether the
gas is heated by starlight or by gravitoturbulence, fragmentation is a
function of $M_{\rm gas}$ (see also Nayakshin 2006 and Levin 2007).

We first consider stellar irradiation, which is dominated by the P1/P2
disk.  The stars in the P1/P2 disk are on average $R_D \sim 1-3$ pc
from any patch of the gaseous disk, whose vertical thickness is $\ll
h$, the thickness of the stellar disk.  For the purposes of this
section, we will adopt fiducial values of $R_D \sim 2$ pc and $h/R_D
\sim 0.3$ motivated by our previous discussion in \S~2.  The flux of
starlight incident on the disk is $F_{*} \sim (L_{*}/4\pi
R_D^2)(h/R_D) \approx 10\, (L_{*}/3\times 10^6\,{\rm L}_{\odot})
(R_D/2\,{\rm pc})^{-2} ([h/R_D]/0.3)\,{\rm ergs\,cm^{-2}s^{-1}}$,
where $L_{*}$ is the total stellar luminosity of the P1/P2 disk.  The
disk is easily optically thick to incident starlight for typical
optical opacities (for dust-to-gas ratios of 0.01; Draine 2003).  We
define the effective temperature from starlight heating as
\begin{equation}\label{eq:Testar}
  \sigma T_{\rm e,*}^4 = F_{*} = \frac {L_{*}}{4\pi R_D^2} \frac h {R_D}, 
\end{equation}
which gives
\begin{equation}
T_{\rm e,*} \approx 20\,\left(\frac {L_{*}}{3\times 10^6\,{\rm
L}_{\odot}}\right)^{1/4}\left(\frac {R_D} {2\,{\rm pc}}\right)^{-1/2}
\left(\frac{h/R_D}{0.3}\right)^{1/4}\ {\rm K}.
\end{equation}
The emitted flux is $F_{\rm IR} \approx \sigma_B T^4 \min(\tau_{\rm
  IR}, 1)$. Equating $F_*$ with $F_{\rm IR}$, we find
\begin{equation}\label{eq:temperatures}
  T = T_{\rm e,*}\min(\tau_{\rm IR}, 1)^{-1/4}\,{\rm K}.
\end{equation}
Note that in the optically thick case, external irradiation generates
a midplane temperature, $T$, that is independent of optical depth. At such
low temperatures (tens of K), the main source of opacity is dust, for
which
\begin{equation}\label{eq:opacity}
  \kappa_{\rm IR} = 
       5.5 \left(\frac {T}{166\,{\rm K}}\right)^2 \,{\rm cm^2\,g^{-1}} \ \ \ \ T
      < 166\,{\rm K}
 \end{equation}
 (Bell \& Lin 1994; Thompson, Quataert, \& Murray 2005).  The disk
 builds up sufficient mass to fragment when $Q\sim 1$, which implies
\begin{equation}\label{eq:Q1critical}
\frac {M_{\rm gas,\,crit}} {M_{\rm BH}} \sim \frac {c_s} {v_{\rm orb}}.
\end{equation}
At this time, the disk may be optically thin or thick.  

We first consider the optically thin case. Combining equations
(\ref{eq:temperatures}), (\ref{eq:opacity}), and
(\ref{eq:Q1critical}), and using $\Sigma_g = M_{\rm gas,crit}/\pi
R^2$ for a gas disk with radius $R$, we find
\begin{eqnarray}
  M_{\rm gas,\,crit} \sim
  5\times 10^4 \left(\frac {M_{\rm BH}} {10^8\Msun}\right)^{6/13} \left(\frac {R}{1\,{\rm pc}}\right)^{8/13} \nonumber \\
\left(\frac {T_{\rm e, *}}{20\,{\rm K}}\right)^{4/13}\,M_{\odot}.
\label{eq:diskQ=1 with stars}
\end{eqnarray}
From equation (\ref{eq:temperatures}), the corresponding temperature is  
\begin{equation}
  T \sim 30 \left(\frac {M_{\rm BH}} {10^8\Msun}\right)^{-1/13} \left(\frac {R}{1\,{\rm pc}}\right)^{3/13} 
\left(\frac {T_{\rm e, *}}{20\,{\rm K}}\right)^{8/13}\,{\rm K},
\end{equation}
and the cooling time is
\begin{equation}
t_{\rm cool} \approx 6 \left(\frac {M_{\rm BH}}
{10^8\Msun}\right)^{5/13} \left(\frac {R}{1\,{\rm
pc}}\right)^{-15/13} 
\left(\frac {T_{\rm e, *}}{20\,{\rm K}}\right)^{-40/13}\,{\rm  yrs}.
\end{equation} 
The cooling condition (eq.[\ref{eq:cooling}]) is satisfied for 
$R \gtrsim R_{\rm cool,thin}$ where
\begin{equation}
R_{\rm cool,thin} \approx 0.1 \left(\frac {M_{\rm BH}} {10^8\Msun}\right)^{23/69} 
\left(\frac {T_{\rm e, *}}{20\,{\rm K}}\right)^{-80/69}\ {\rm pc}.
\end{equation}
Once the critical gas mass is reached (eq.[\ref{eq:diskQ=1 with
  stars}]) (for $R \gtrsim R_{\rm cool,thin}$), the disk
fragments and forms stars.

For larger gas masses, the disk becomes optically thick.  Using
equations (\ref{eq:opacity}) and (\ref{eq:Q1critical}), we find that
the gas mass where the optically thin to thick transition occurs is
\begin{equation}
  M_{\rm gas,\tau = 1} \approx 6\times 10^4 \left(\frac {R} {1\,{\rm pc}}\right)^{4/5} \left(\frac {M_{\rm BH}} {10^8\Msun}\right)^{-2/5} \Msun.
\end{equation}
For an optically thick disk, the critical mass for fragmentation is
\begin{eqnarray}
  M_{\rm gas,\,crit} \sim 3\times 10^4 \left(\frac {M_{\rm BH}} {10^8\Msun}\right)^{1/2} \left(\frac {R}{0.3\,{\rm pc}}\right)^{1/2} 
  \nonumber \\ \left(\frac {T_{\rm e, *}}{20\,{\rm K}}\right)^{1/2}M_{\odot}\label{eq:Mcritthick},
\end{eqnarray}
where the corresponding temperature $T=T_{\rm e,*}$.  Note we rescaled
$T_{\rm e,*}$ so that the disk is self-consistently optically thick.
The cooling time is
\begin{equation}\label{eq:tcoolthick}
t_{\rm cool} \approx 100 \left(\frac {R} {0.3\,{\rm
pc}}\right)^{-3}\left(\frac {M_{\rm BH}}{10^8\Msun}\right)\,{\rm yrs}.
\end{equation}
For optically thick cooling in the regime where $\kappa \propto T^2$,
the cooling time is independent of the gas mass.  In this case, the
cooling condition (eq.[\ref{eq:cooling}]) is satisfied for $R
\gtrsim R_{\rm cool,thick} = 0.2 (M_{\rm BH}/10^8\Msun)^{-1/3}$ pc.


If a disk reaches $Q\sim 1$, but the cooling condition
(eq.[\ref{eq:cooling}]) is not satisfied, then it cannot fragment
immediately.  Instead, gravitoturbulence heats the disk to maintain a
temperature of
\begin{equation}\label{eq:diskTQ=1}
  T \approx 30 \,  \left(\frac {R} {1\,{\rm pc}}\right)^{-1} \left(\frac {M_{\rm gas}}{5\times 10^4\Msun}\right)^2 \left(\frac {M_{\rm BH}} {10^8\Msun}\right)^{-1}\,{\rm K}
\end{equation}
so as to keep $Q\sim 1$.  To fragment, the disk must accumulate
additional mass until $\Omega t_{\rm cool} \lesssim 3$. Again this
cooling may proceed via optically thin or thick emission.  The
critical mass for fragmentation for an optically thin disk is
\begin{equation}\label{eq:disk fragmentation mass}
  M_{\rm gas, \,crit} \approx 3\times 10^4 \left(\frac {M_{\rm BH}} {10^8\Msun}\right)^{11/20} \left(\frac {R} {1\,{\rm pc}}\right)^{7/20}\Msun.
\end{equation}
For $R < R_{\rm cool,thick}$, optically thick cooling in the $\kappa
\propto T^2$ regime is too slow for the disk to fragment.  Instead,
the gas mass there will build up and gravitoturbulence will heat the
disk until $\kappa$ is not longer proportional to $ T^2$, i.e., $T >
166\,{\rm K}$.  Above this temperature, $\kappa \approx 5.5\,{\rm
  cm^2\,g}^{-1}$ is roughly independent of temperature (c.f. Thompson
et al. 2005), though it varies by factors of a few between $T\sim 100
- 1000\,{\rm K}$, where the upper bound is set by the dust sublimation
temperature.  Assuming a constant opacity, we find the critical gas
mass for fragmentation to be
\begin{equation}\label{eq:tcoolconstant}
M_{\rm gas,\,crit} \approx 5 \times 10^4 \left(\frac {M_{\rm
BH}}{10^8\Msun}\right)^{7/8}\left(\frac{R}{0.2\,{\rm
pc}}\right)^{-5/8}\Msun,
\end{equation}
for $R < R_{\rm cool,thick}$.

\begin{figure}
\plotone{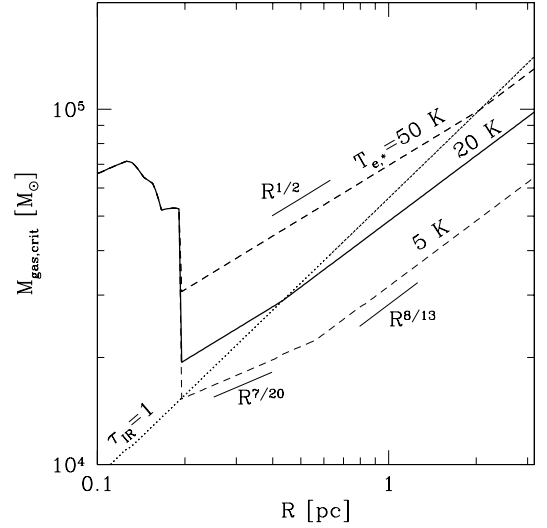}
\caption{Critical gas mass for fragmentation as a function of disk
  radius $R$ at $T_{\rm e,*}=5$ (lower dashed line), $20$ (thick solid
  line), and $50$ K (upper dashed line) for $M_{\rm BH} = 10^8\Msun$.
  At $R \lesssim 0.2$ pc, local accretion heating dominates, while at
  larger radii, irradiation dominates.  The dotted line shows the disk mass
  at which $\tau_{\rm IR} = 1$. We also show the $R^{8/13}$ scaling from equation
  (\ref{eq:diskQ=1 with stars}), the $R^{1/2}$ scaling from equation
  (\ref{eq:Mcritthick}), and the $R^{7/20}$ scaling from equation
  (\ref{eq:disk fragmentation mass}).}
\label{fig-Mdiskcrit}
\end{figure}

We summarize the above results in Figure \ref{fig-Mdiskcrit}.
We use the opacity table compiled by Semenov et al.~(2003) and
numerically compute the self-consistent fragmentation mass
(eqns.[\ref{eq:Toomre}] and [\ref{eq:cooling}]) with radiative cooling
(eq.[\ref{eq:coolingtime}]), allowing for starlight
(eq.[\ref{eq:temperatures}]) and gravitoturbulent heating
(eq.[\ref{eq:diskTQ=1}]), whichever is greater. 
Figure \ref{fig-Mdiskcrit} shows the critical disk mass for
gravitational collapse as a function of $R$ at $T_{\rm e,*}=5$,
$20$, and $50$ K.  
We also show the $R^{8/13}$ scaling from equation (\ref{eq:diskQ=1 with stars}), the
$R^{1/2}$ scaling from equation (\ref{eq:Mcritthick}), and the
$R^{7/20}$ scaling from equation (\ref{eq:disk fragmentation
  mass}) in their respective regimes.  We find that at $R
\approx 1$ pc, external irradiation dominates.  For $R\lesssim
R_{\rm cool,thick} \approx 0.2$ pc, gravitoturbulence heats the
central disk temperature above which the opacity law, $\kappa \propto
T^2$, no longer holds.  However, we do not recover the scaling
suggested by equation (\ref{eq:tcoolconstant}) as the opacity is not a
constant above $166\,{\rm K}$, but rather varies by factors of a few.

We have shown that depending on whether disks are externally heated
(eq.  [\ref{eq:diskQ=1 with stars}]) or internally heated (eq.
[\ref{eq:disk fragmentation mass}]), the fragmentation mass is
$3-5\times 10^4\Msun$.  Gas from stellar mass loss at $r > R_{\rm t}$
is driven to $r \approx R_{\rm t}$ on a timescale $t_{\rm flow} \sim
10^6$ yrs (\S 2). For mass supply rates of $\dot{M}_*\sim 10^{-4}
\Msun \, {\rm yr^{-1}}$, the steady-state disk mass outside $R_{\rm
  t}$ is expected to be $\sim 100 \Msun$, well below that required to
fragment.  Thus all of the gas is driven to $R \approx R_{\rm t}
\sim 1$ pc, where it collects in a ring.  
The timescale for gas to viscously spread once it accumulates at $R_{\rm t}$ is 
\begin{eqnarray}
  t_{\rm visc} = \frac {t_{\rm dyn}} {\alpha} \left(\frac h {R_{\rm t}}\right)^{-2}  
  \approx 5000 \alpha^{-1} \left(\frac {M_{\rm BH}} {10^8\Msun}\right)^{1/2} \nonumber \\ \left(\frac{R_{\rm t}}{1\,{\rm pc}}\right)^{1/2}\left(\frac T {30\,{\rm K}}\right)^{-1}\,{\rm Myrs}, 
\end{eqnarray}
where $\alpha < 1$ is the dimensionless viscosity. We compare this to
the time needed to accumulate a gravitationally unstable disk, $t_{\rm
  accum} = M_{\rm gas,crit}/\dot{M}_*$.  The mass loss rate varies as
the stellar population ages.  Using Starburst99, we find that as a
$2\times 10^7\Msun$ stellar cluster ages from $3\times 10^9$ to $10^{10}$ yrs,
the mass loss rate, $\dot{M}_*$, ranges from $3\times 10^{-4}$ to
$3\times 10^{-5}\Msun\,{\rm yr}^{-1}$.  The range in $\dot{M}_*$ and
the range in $M_{\rm gas,crit}$ yield $t_{\rm accum} \approx 100 - 2000$
Myrs.  Hence the ratio
\begin{equation}
  \frac {t_{\rm accum}}{t_{\rm visc}} \approx  0.1 
  \alpha\left(\frac {\dot{M}_*}{10^{-4}\Msun\,{\rm yr}^{-1}}\right)^{-1} \left(\frac T {30\,{\rm K}}\right)^{3/2},
\end{equation}
where we have used the critical mass for $Q\sim 1$ from equation
(\ref{eq:Q1critical}).  Thus, even for $\alpha \sim 1$, the ratio
$t_{\rm accum}/t_{\rm visc} \lesssim 0.3$, which implies that
gas fragments at $\sim R_{\rm t}$ before accreting.


The current mass in stars in P3 is estimated to be $\approx 4200
\Msun$ (B05).  The disk mass required for gravitational instability is
$\sim 6-10$ times higher than this (Fig. \ref{fig-Mdiskcrit}).  Such
a difference suggests a star formation efficiency $\sim 10-20\%$.  A
larger star formation efficiency can be accommodated if the initial
mass function (IMF) of stars in this extreme environment is top heavy,
which would help reconcile our estimated $M_{\rm gas,crit}$ with the
current inferred mass of P3.  There is theoretical (Nayakshin 2006;
Levin 2007) and observational (Nayakshin \& Sunyaev 2005; Paumard et
al. 2006) evidence for a top heavy IMF in the GC starburst.

Because $t_{\rm dyn} \ll t_{\rm accum}$, it is also possible that only
a small fraction of the gas fragments into stars once the disk mass
exceeds $M_{\rm gas,crit}$, leaving behind a stable disk with a mass
only modestly below $M_{\rm gas,\,crit}$.  The excess gas would remain
in the nuclear region, given the severe difficulty that stellar winds
and supernovae would have in removing it from so deep in the
gravitational potential of the black hole.  In this case, the gas disk
in the nucleus of M31 would maintain a mass of $\sim M_{\rm
  gas,\,crit}$.

Since the starburst that produced P3 occurred, we expect
$\dot{M}_* \times 200\,{\rm Myr}\sim 10^4\Msun$ of gas to have
accumulated near $R_{\rm t}\sim 1$ pc from ongoing stellar mass
loss. This molecular gas would be analogous to the circumnuclear disk
(CND) in the GC, with temperatures of $T\approx 30$ K and extremely
high densities of $n\sim 10^{9}\,{\rm cm}^{-3}$ ($Q\gtrsim 1$).  This
gas would be bright in CO and HCN; the CO (1-0) flux would be $\approx
2$ mJy (for an optically thick line).

%

\section{Discussion and Conclusions}\label{sec:conclusions}

We have argued that the origin of the young stars in P3 in M31 is
rooted in the P1/P2 disk. The non-axisymmetric component of
the potential from the P1/P2 disk restricts non-intersecting gas
orbits to distances $r\lesssim R_{\rm t} \sim 1$ pc from the
central SMBH if the pattern speed (i.e., precession frequency) of the
P1/P2 disk is $\Omega_p \lesssim 3-10\,{\rm km\,s^{-1}pc^{-1}}$.  At
larger radii, gas finds itself in intersecting orbits; it shocks,
loses energy, and gets driven to $R_{\rm t} \sim 1$ pc.  This is
comparable to the maximum radial extent of the A stars of P3.

Stellar mass loss from the P1/P2 disk can supply the gas that formed
P3.  Stellar winds supply mass at a rate of $\sim 10^{-4}\Msun\,{\rm
  yr}^{-1}$ for a $\sim 10^{10}$ yr old population.  This gas
accumulates in a disk at $r\lesssim R_{\rm t}$. The conditions for
fragmentation (eqns.[\ref{eq:Toomre}] and [\ref{eq:cooling}]) give a
critical gas mass of $\sim 5\times 10^4\Msun$.  Hence, every $\sim
500$ Myr, the disk accumulates enough mass to fragment and produces a
starburst.  This recurrence time is consistent with the age of the A
stars of 200 Myr (B05).  In addition, the observed alignment of the P3
disk with the P1/P2 disk is consistent with our argument that the
P1/P2 disk supplies the gas which formed P3.

Several predictions arise naturally out of this model.  First, the
pattern speed of the eccentric stellar disk should be $\Omega_p
\lesssim 3-10\,{\rm km\,s^{-1}pc^{-1}}$.  Current observational
constraints on the pattern speed are weak (see Appendix A).  Second,
there should be $\sim 10^4\Msun $ of gas in the nucleus of M31 from
accumulation of gas since the last starburst that produced P3.  This
molecular gas would be analogous to the CND at the GC.  Such gas would
have a temperature of $\sim 30$ K, be at extremely high densities
$n\sim 10^{9}\,{\rm cm}^{-3}$, and be bright in CO and HCN with a CO
(1-0) flux of $\approx 2$ mJy and a line width of $\approx 1000\,{\rm
  km\,s}^{-1}$.  Under these conditions, dust will emit with a flux of
$\sim 20$ mJy at 70 $\micron$, $\sim 100$ mJy at 160 $\micron$, and
$\sim 0.1$ mJy at 1 mm.  Finally, starbursts at $\lesssim 1$ pc in M31
should occur every $10^8-10^9$ yrs.  These older generations of stars
and stellar remnants may be detectable.

Another interesting possibility, which was suggested by the referee,
is that the P3 stars may be chemically anomalous because they are
constructed from the recycled AGB winds of the P1/P2 disk.  A similar
self-pollution scenario has been proposed in the context of globular
clusters (Cottrell \& Da Costa 1981; Ventura et al. 2001).  The
composition of the AGB wind depends strongly on the initial mass of
the star and the amount of dredge-up between the core and the envelope
(Fenner et al.  2004), which is uncertain.  In light of these
uncertainties, it is interesting to note that HST observations of
omega Centauri, the largest Galactic globular cluster.  show a double
main sequence (Bedin et al. 2004), suggesting two episodes of star
formation.  The blue main sequence, which arises from the second
episode of star formation, may possess a considerable enhancement of
helium (Bekki \& Norris 2006; Karakas et al. 2006), which could come
from the AGB winds of the red main sequence, i.e., the stars of the
first episode of star formation.  It would be interesting to search
for an analogous chemical anomaly in the P3 stars of M31.




Observations indicate that P3 appears to be a circular disk around the
SMBH. Gas orbits around the SMBH are eccentric in the presence of the
P1/P2 disk as illustrated in Figure \ref{fig:3km}. However, stars that
form from an eccentric gas disk may not have an eccentric distribution
themselves at later times. Once the $Q\sim 1$ gas disk turns into
stars, these stars will precess differentially because of their own
self-gravity.  We estimate the differential precession rate to be
\begin{equation}\label{eqn-precess}
\frac{d\dot\varpi}{dR}\Delta R \sim \Omega \left(\frac{M_{\rm P3}}{M_{\rm BH}}\right)\left(\frac{\Delta R}{R}\right)
\end{equation}
where $\dot\varpi$ is the precession frequency of a star at radius $R$, and
$\Delta R$ is the initial radial extent of the P3 stellar disk with mass
$M_{\rm P3}$.  A
spread of $\Delta R/R \sim 0.1$ in the orbits of the P3 stars can be
generated by viscous spreading of the deposited gas prior to star
formation.
Taking ${M_{\rm P3}}/{M_{\rm BH}} \sim 10^{-4}$, we find that stars in
P3 differentially precess out of their initially apsidally aligned
structure over a timescale $10^{5} \Omega^{-1} \sim 10^8$ years,
comparable to the age of the A stars. 

Over 10 Gyrs, a mass loss rate of $3\times 10^{-5}$ to $3\times
10^{-4}\Msun\,{\rm yr}^{-1}$ from the P1/P2 disk will redistribute
$\approx 3\times 10^5 - 3\times 10^6\Msun$ of mass to the P3 disk,
which is of order 10\% the mass of the P1/P2 disk.  If a large
fraction of this mass is retained, it may affect the eccentricity of
the P1/P2 disk.  The backreaction of a growing P3 disk on the
eccentric P1/P2 disk is beyond the scope of this paper, but is an
interesting question for further study.

Our model may be applicable to other galaxies with double nuclei in
addition to M31.  Lauer et al.  (1996) observed a double nucleus in
NGC 4486B with a 12 pc separation.  Debattista et al. (2006) detected
a double nucleus in VCC 128 with a 32 pc separation.  Thatte et al.
(2000) also detected a double nucleus in the starburst galaxy M83 with
a 5.4 pc separation.  If these double nuclei are associated with
non-axisymmetric stellar distributions, very compact nuclear
starbursts and dense nuclear molecular gas may be common features of
galactic nuclei.

Finally, we briefly discuss our model in the context of the GC.
Observations suggest that the $1.3\times 10^4\Msun$ of young massive
stars in the GC are concentrated between $r\sim0.04-0.4$ pc (Ghez et
al. 2005; Paumard et al. 2006), similar in mass and radial extent to
M31.  A non-axisymmetric component of the potential may explain the
radial extent of these young stars, which otherwise can only be
accounted for by the assumption that gas is supplied on very low
angular momentum orbits.  If the non-axisymmetric component were due
to an eccentric disk of old stars, it would likely remain undetected
because of extinction.  



\acknowledgements

We thank R. Genzel, A. Loeb, B. Paczynski, L. Strubbe, and S. Tremaine
for useful discussions.  We thank L. Blitz and A. Bolatto for useful
discussions and for performing CO observations on M31.  We thank the
anonymous referee for useful comments.  We thank B.  Johnson for
presenting a talk on M31 in the Astro Reading Group (ARG) that led to
this project. We also thank G.  Bower, G. Howes, B.  Metzger, and B.
Schmekel for leading other seminars in the ARG.  We would also like to
acknowledge all ARG participants.  P.C. thanks the Institute for
Advanced Study and the Canadian Institute for Theoretical Astrophysics
for their hospitality.  P.C. is supported by the Miller Institute for
Basic Research. R.M.-C. is supported by a NSF graduate fellowship.
E.C.  is supported in part by an Alfred P. Sloan Fellowship and
NSF-AST grant 0507805.  E.Q. was supported in part by NASA grant
NNG05GO22H and the David and Lucile Packard Foundation.


\appendix
\section{Dynamics of P1 and P2}\label{sec:PT}

The most widely accepted model for the double nucleus of M31 is that
of an eccentric stellar disk (T95).  T95 fits the light distribution
with three apsidally aligned ellipses with the SMBH at one focus.
Stars on these elliptical Keplerian orbits pile up at apoapse, giving
rise to P1.  A suitable nesting of orbits gives rise to P2 at
periapse.  PT03 performed a more careful calculation to fit the light
and velocity distributions, using Keplerian ellipses again.  These
models do not take disk self-gravity into account.  Since the stellar
disk mass is nearly $2\times 10^7\Msun$, comparable to the SMBH mass
of $\sim 10^8\Msun$ (B05), disk self-gravity has significant effects
on the dynamics of both stars and gas.

Other workers have developed self-consistent models for the dynamics
of such an eccentric stellar disk.  Statler (1999) created
self-consistent models for a stellar disk assuming that stars follow
periodic orbits.  This work was expanded by Salow \& Statler (2004),
who find high pattern speeds $\Omega_p \sim 30\,{\rm
  km\,s^{-1}\,pc^{-1}}$.  Sambhus \& Sridhar (2002) used families of
prograde and retrograde loop orbits to fit the light distribution and
found $\Omega_p \approx 16\,{\rm km\,s}^{-1}$.  Bacon et al. (2001)
showed through N-body simulations that the $m=1$ perturbation can be
long-lived, $\sim 100$ Myrs for low pattern speeds $\Omega_p \lesssim
3\,{\rm km\,s^{-1}\,pc^{-1}}$.  Similarly, Jacobs \& Sellwood (2001)
found long-lived $m=1$ modes in an eccentric ring with a pattern speed
of $\Omega_p/\Omega_0 \approx 0.4 M_D/M_{\rm BH}$, where $\Omega_0 =
\sqrt{GM/R_0^3}$ and $R_0 \approx 2$ pc.  This corresponds to a
pattern speed of $\Omega_p \approx 16\,{\rm km\,s^{-1}\,pc^{-1}}$ for
the parameters appropriate to M31. Finally, Sambhus \& Sridhar (2000)
utilized a variant of the Tremaine-Weinberg method (Tremaine \&
Weinberg 1984) to obtain an observational constraint on the pattern
speed of $<30\,{\rm km\,s^{-1}\,pc^{-1}}$ (for a disk inclination
relative to the sky plane of $77$ degrees).

As the constraints on the pattern speed are weak, we have elected to
take PT03's fits to the P1/P2 disk and to experiment with a range of
values of $\Omega_p$.  We use their formulae for the disk eccentricity
profile and surface density (their eqs.[12] and [17]):
\begin{eqnarray}
e_m(a) = \alpha\left(a_e - a\right)\exp\left[-\frac{\left(a - a_g\right)^2}{2w^2}\right], \\
\Sigma(a) = \Sigma_0\frac{a^2\exp\left(-a/a_0\right)}{1 + \exp\left[4\left(a - c_2\right)\right]},
\end{eqnarray}
where $a$ is the semi-major axis in pc, $e_m$ is the mean
eccentricity, and the effective surface density, $\Sigma(a)$, is
defined as $dM = 2\pi a\Sigma(a)da$, where $dM$ is the mass between
$a$ and $a+da$.  We take the maximum semi-major axis to be $a = 8$ pc.
The normalization $\Sigma_0$ is set by $M_D/M_{\rm BH}$, which we
vary.  For convenience we list the fitting parameters (their Table 2)
in our Table 1.  We focus on PT03's nonaligned model in this paper,
but we have studied their aligned model as well.  In the aligned
model, for $R_{\rm t} \lesssim 1$ pc, we find the required pattern
speed is $\Omega_p\lesssim 6\,{\rm km\,s^{-1}pc^{-1}}$ for $M_D/M_{\rm
  BH} = 0.1$ and $M_{\rm BH} = 10^8\Msun$ for the softening length of
$h=0.1$ pc, similar to the nonaligned model.




 
\begin{deluxetable}{c c c}
  \tablecolumns{6}
  \tablewidth{0pt}
  \tablecaption{Model parameters from Peiris \& Tremaine (2003).
    \label{table:parameters}} 

  \tablehead{\colhead{Parameter} & \colhead{Aligned Model}
    & \colhead{Unaligned Model} }
  
  \startdata
  $\alpha$ & 0.288 & 0.197 \\
  $a_e$ [pc] & 3.97 & 4.45 \\
  $a_g$ [pc] & 1.51 & 1.71 \\
  $w$ [pc] & 1.53 & 1.52 \\
  $a_0$ [pc] & 4.61 & 1.37 \\
  $c_2$ [pc] & 3.79 & 4.24 
  \enddata

\end{deluxetable}

\end{document}